\documentclass[twocolumn,showpacs,preprintnumbers,amsmath,amssymb]{revtex4}

\usepackage{graphicx}
\usepackage{dcolumn}
\usepackage{bm}

\begin{document}

\title{Josephson-vortex-flow terahertz emission in layered high-$T_c$ superconducting single crystals}


\author{Myung-Ho Bae$^1$}
\author{ Hu-Jong Lee$^{1,2}$}
\author{Jae-Hyun Choi$^1$}

\affiliation{$^1$Department of Physics, Pohang University of
Science
and Technology, Pohang 790-784, Republic of Korea}%
\affiliation{$^2$National Center for Nanomaterials Technology,
Pohang 790-784, Republic of Korea}
\date{\today}

\begin{abstract}
We report on the successful terahertz emission (0.6$\sim$1 THz)
that is continuous and tunable in its frequency and power, by
driving Josephson vortices in resonance with the collective
standing Josephson plasma modes excited in stacked
Bi$_2$Sr$_2$CaCu$_2$O$_{8+x}$ intrinsic Josephson junctions.
Shapiro-step detection was employed to confirm the terahertz-wave
emission. Our results provide a strong feasibility of developing
long-sought solid-state terahertz-wave emission devices.
\end{abstract}


\maketitle

Rapidly increasing applications of the electromagnetic (EM) waves
in the terahertz frequency range ($\sim$0.1-30 THz) to
nondestructive diagnosis, safe medical imaging, security measures
and inspections, high-speed communications, etc., demand a novel
technique of efficient terahertz-wave generation. Semiconductor
EM-wave generating elements, covering a wide-range of frequency
spectra, mainly rely on two operation principles: sinusoidal
current oscillation and the transition of electronic states
between quantum levels, which are realized by electronic and
photonic means, respectively. However, terahertz-range EM-wave
technology has remained underdeveloped due to the following
reasons: (i) in electronics, the transit of charge carriers in a
semiconductor device inherently takes too long to produce the
terahertz oscillation, while (ii) in photonics, the
terahertz-range photon energy is much lower than the
room-temperature ambient thermal energy. This lack of generating
technology in the terahertz band is often dubbed the ``terahertz
gap". In spite of recent progress in assorted generation
techniques \cite{Ferguson, Kohler, Carr, Kawase} the terahertz
generation remains to be a subject of intensive research efforts.

The present work describes the direct detection of the Josephson
vortex-flow-induced terahertz-wave emission from
Bi$_2$Sr$_2$CaCu$_2$O$_{8+x}$ (Bi-2212) stacked intrinsic
Josephson junctions (IJJs). Since the size of the superconducting
gap in Bi-2212 is higher than the Josephson plasma edge both the
emission and the detection are allowed in a terahertz range
\cite{Singley}. In this study, the terahertz-wave emission was
confirmed by observing Shapiro steps in an on-chip stack of IJJs
(detector stack) as it was irradiated by the emission from the
oscillator stack. The frequency of the emitted wave was tuned by
the bias voltage, while the emission power was controlled by the
bias current and switching of the resonating modes.

Naturally grown layered high-T$_c$ superconducting Bi-2212 single
crystals contain IJJs \cite{Kleiner}, where a 0.3-nm-thick
superconducting CuO$_2$ bi-layer and a 1.2-nm-thick insulating
BiO-SrO layer alternate along the $c$ axis. Since the
superconducting electrodes in IJJs are much thinner than the
$c$-axis London penetration depth $\lambda_{ab}$ ($\sim$200 nm)
the phase parameter $\phi_n$ in the $n$th Josephson junction is
inductively coupled to those of neighboring junctions
\cite{Sakai,Machida}. Thus, the electrodynamics, governed by
coupled Sine-Gordon differential equations in a system of $N$
stacked IJJs, is represented by the same number of standing
Josephson plasma (SJP) collective eigen modes formed along the
$c$-axis direction. Josephson vortices generated by a planar
external magnetic field tend to move in a tunneling bias current
in resonance with the collective SJP modes, with the spatial phase
distribution ranging from a rectangular (in-phase) to a triangular
(out-of-phase) lattice. The Josephson-vortex-induced emission
power can be enhanced by tuning the vortex lattice closer to the
in-phase rectangular mode \cite{Koshelets}.

\begin{figure}[b]
\begin{center}
\leavevmode
\includegraphics[width=1\linewidth]{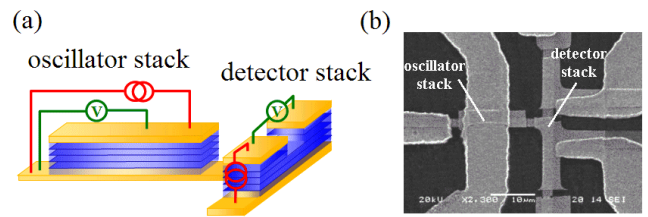}
\caption{(color online) (a) Schematic of sample and measurement
configurations, showing the oscillator stack (left) and the
detector stack (right) of intrinsic Josephson junctions. (b) SEM
image of SP2.}
\end{center}
\end{figure}

The conventional solid-state-reaction method was adopted to
prepare Bi-2212 single crystals, which were slightly overdoped.
The double-side-cleaving technique \cite{Wang} was then employed
to isolate and sandwich a finite number of Bi-2212 IJJs between
two Au electrodes, essentially eliminating the interference from
the pedestals (stacks of IJJs outside of but coupled to the stack
of IJJs of interest) and thus allowing the formation of ideal
multiple SJP modes along the $c$-axis direction. We fabricated two
specimens of stacked IJJs (SP1 and SP2) sandwiched between two Au
electrodes. Detailed fabrication procedures are described in Ref.
\cite{MH}. Figs. 1(a) and (b) show the schematic of the typical
sample configuration and an SEM image of SP2, respectively. The
oscillator stacks were 13.5$\times$1.4 $\mu$m$^2$ [SP1] and
15.2$\times$1.9 $\mu$m$^2$ [SP2], while the detector stacks were
18$\times$1.3 $\mu$m$^2$ [SP1] and 11.6$\times$2.3 $\mu$m$^2$
[SP2] in their lateral dimensions.

\begin{figure}[b]
\begin{center}
\leavevmode
\includegraphics[width=0.8\linewidth]{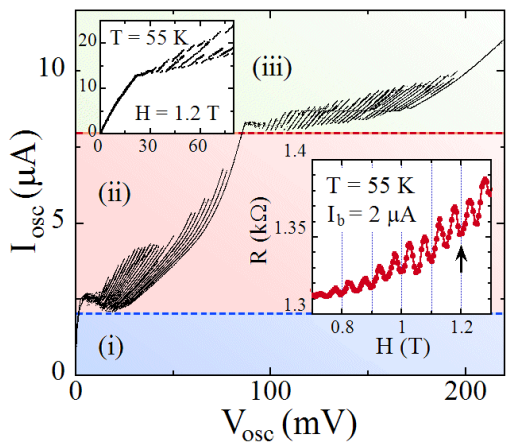}
\caption{(color online) {\it I-V} characteristics corresponding to
the collective Josephson vortex states [(i), (ii), and (iii);
refer to the main text] of SP1 for $H$ =3.9 T at $T$=4.2 K. The
contact resistance was numerically subtracted. Upper inset: a
single vortex-flow branch (low biases) and quasiparticle branches
(high biases) in $H$=1.2 T at $T$=55 K, corresponding to the arrow
position in the lower inset. Lower inset: Josephson vortex-flow
resistance versus external magnetic field of SP1 at $T$=55 K in a
bias current of 2 $\mu$A.}
\end{center}
\end{figure}

The lower inset of Fig. 2 shows the oscillation of a Josephson
vortex-flow resistance (JVFR) of SP1 ($N$=35) as a function of
external magnetic field at $T$=55 K for a bias current of $I_b$=2
$\mu$A. The field period of the oscillation, 0.51 kG, is one half
of $H_0$ [=$\Phi_0/Ld$=1.02 kG]. Here, $H_0$ is the magnetic field
corresponding to a flux quantum, $\Phi_0$, through each junction
of length $L$ (=13.5 $\mu$m) and thickness $d$ (=1.5 nm). The
$H_0$/2 JVFR oscillation, often observed in the S-shaped embedded
structures, is known to originate from the interaction between a
triangular vortex lattice and the edge potential for the vortex
entry and exit at the junction boundaries \cite{Ooi, Machida2}.
The $H_0$/2 oscillation in our sandwiched oscillator stack thus
strongly indicates the formation of the triangular lattice in its
low-bias static limit.

The upper inset of Fig. 2 exhibits a single vortex-flow branch
corresponding to the near-static triangular lattice in the
low-bias region (V$<$23 mV) and multiple quasiparticle branches in
high biases at $T$=55 K in a transverse field of $H$=1.2 T; the
field denoted by the arrow in the lower inset. In this field
range, the single vortex-flow branch persists down to $T$=4.2 K
(not shown). In a higher magnetic field of $H$=3.9 T as in the
region (ii) of the main panel of Fig. 2, however, the single
vortex-flow branch splits into multiple vortex-flow sub-branches,
which arise as a result of phase-locking of a Josephson vortex
lattice on multiple SJP modes \cite{Bae,Lee}. SJP oscillations in
each mode, standing along the $c$ axis, propagate along junctions
with a characteristic collective mode velocity
\cite{Sakai,Machida}, $c_n=c_0/\sqrt{1-\mbox{cos}[\pi n/(N+1)]}$.
Here, $n$ (=1$\sim$$ N$) is the mode index and $c_0$ is the
Swihart velocity, which governs the EM-wave propagation in a
single Josephson junction.

In SP1 the JVFR oscillations are limited in the current bias range
up to $I_b$$\sim$2 $\mu$A. The multiple vortex-flow sub-branches
in the bias range of 2 $\mu$A$<I_b<$8 $\mu$A are clearly
distinguished from the multiple quasiparticle branches above
$I_b$$\sim$8 $\mu$A, which corresponds to the critical current for
the Josephson pair tunneling in the given magnetic field. In
general, a stack of IJJs driven by a bias current in a planar
magnetic field is in one of the three different dynamic
Josephson-vortex states as illustrated in Figs. 2 and 3(a): (i)
the near-static triangular vortex configuration, (ii) the
multiple-mode dynamic vortex configuration, and (iii) the
irregular vortex configuration corresponding to the McCumber state
dominated by the quasiparticle tunneling.

The maximum vortex-lattice velocity is governed by the propagation
velocity of a resonating SJP mode in a stack of Josephson
junctions. Thus, for the bias beyond $I_b$$\sim$8 $\mu$A, the
vortex-flow state transits to the quasiparticle tunneling state as
the flowing vortices pass the state of the maximally allowed
cut-off voltage, $V_{max}$, set by the highest plasma velocity in
a junction. The maximum vortex-lattice velocity, $c_{max}$, in SP1
was estimated with $V_{max}$ to be $\sim$3.8$\times$10$^5$ m/s
using the relation \cite{Bae} of $V_{max}$=$NHdc_{max}$. The value
of $c_{max}$ in this sandwiched oscillator stack should represent
the fastest propagation mode among $N$ collective SJP modes. The
Swihart velocity estimated based on the resistively and
capacitively shunted junction model \cite{Bae,Irie} turns out to
be $c_0$=2.0$\times$10$^4$ m/s. Then the corresponding highest
in-phase mode velocity, $c_1$=3.5$\times$10$^5$ m/s, is in
remarkable agreement with the observed value of $c_{max}$ in SP1.
This strongly implies that the last sub-branch [the rightmost
branch in Region (ii)] corresponds to the in-phase rectangular
vortex lattice.

The moving vortex lattice that is in resonance with the collective
transverse SJP oscillations is theoretically predicted to emit
highly coherent EM waves at a junction edge
\cite{Machida,Tachiki}. The detection of the Shapiro steps, which
arise as a result of the resonance between supercurrent
oscillations and an irradiated microwave in a Josephson junction
\cite{Doh}, would be the most direct confirmation of the
electromagnetic emission from the sandwiched oscillator stack,
with information on both the frequency and the power of the
emission.

\begin{figure}[b]
\begin{center}
\leavevmode
\includegraphics[width=0.65\linewidth]{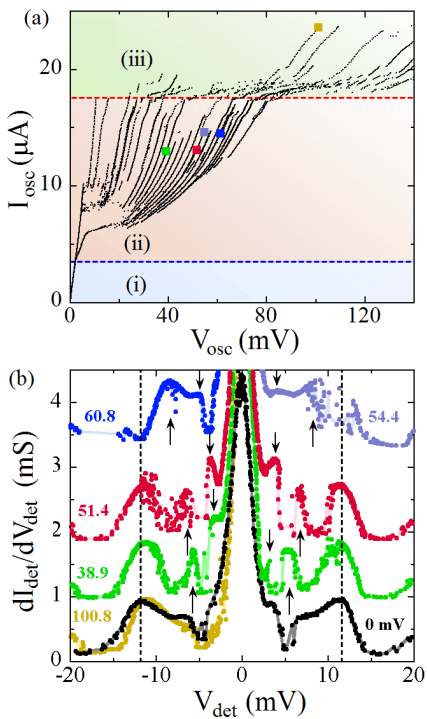}
\caption{(color online) (a) Collective Josephson-vortex-flow
multiple sub-branches and bias voltages $V_{osc}$ (filled squares)
of the oscillator stack of SP2 for $H$=4 T at $T$=4.2 K. The
contact resistance was numerically subtracted. The two horizontal
dashed lines divide the {\it I-V} characteristics into three
regions as in Fig. 2. (b) Responses of the detector stack revealed
in its differential conductance corresponding to the biases of the
oscillator in (a). Each curve is shifted vertically for clarity.}
\end{center}
\end{figure}

Fig. 3(a) shows the bias conditions of the oscillator stack of SP2
($N$=28), $V_{osc}$=38.9, 51.4, 54.4, and 60.8 mV (denoted by
filled squares in its vortex-flow branches in the dynamic vortex
state) and 100.8 mV (denoted in the quasiparticle branch in the
McCumber state), in a field of 4 T at $T$=4.2 K. In SP2 the
detector stack was coupled by the bottom Au electrode with the
2-$\mu$m interspacing to the oscillator stack.

The black $dI_{det}/dV_{det}$-vs-$V_{det}$ curve at the bottom of
Fig. 3(b) represents the response of the detector stack without
any bias to the oscillator stack. The curve corresponds to simple
nonlinear {\it I-V} characteristics without sub-branches in the
vortex-flow region (not shown). The pronounced enhancement of the
zero-bias conductance is due to Josephson vortex pinning. The
vertical dotted lines indicate the positions of $V_{max}$ of the
detector stack, where the broad humps are generated by the change
in the slope of the detector {\it I-V} curves at the boundary
between the vortex-flow and quasiparticle tunneling regions. In
order to reduce the vortex entry to the detector stack, its width
facing the magnetic field (2.3 $\mu$m) was designed to be much
smaller than that of the oscillator stack. In consequence, not all
the IJJs in the detector stack took part in the vortex flow
motion. The effective number of vortex-flowing junctions in the
detector was estimated to be four by comparing values of $V_{max}$
between the oscillator and detector stacks \cite{Wang2}. Here, we
assumed the maximum velocity of the vortices was same in both
oscillator and detector stacks \cite{Carapella}. For a selected
fixed bias to the oscillator stack in Fig. 3(a), two clear
conductance peaks appear in the corresponding
$dI_{det}/dV_{det}$-vs-$V_{det}$ curve of the detector stack:
$V_{det}^{prim}(V_{det}^{sub})$=5.3$\pm$0.5 (2.8$\pm$0),
6.6$\pm$0.2 (3.7$\pm$0.2), 7.9$\pm$0.3 (4.1$\pm$0.2), and
8.5$\pm$0.1 (5.2$\pm$0.4) mV for four bias conditions in order of
increasing voltages in the dynamic vortex state, respectively.
Here, $V_{det}^{prim}$ and $V_{det}^{sub}$ are the voltage
positions of the primary peak and the sub-peak, which are denoted
by the upward and the downward arrows, respectively, in Fig. 3(b).
The primary conductance peaks are always more evident than the
sub-peaks. One also notes that the voltage values of peaks in each
curve shift in proportion to the increase of the bias voltage
$V_{osc}$ of the oscillator stack.

To confirm that the observed peaks are indeed the response of the
detector stack to the irradiation \cite{Ustinov}, the voltages of
the conductance peaks ($v_{det}$) are plotted in Fig. 4(a) as a
function of the bias voltage $v_{osc}$. Here, $v_{osc}$ ($\equiv
V_{osc}$/$N_{osc}$) and $v_{det}$ ($\equiv V_{det}$/$N_{det}$) are
normalized by the number of junctions involved in the vortex flow
in the oscillator and the detector stacks ($N_{osc}$=28 and
$N_{det}$=4), respectively. For the primary peaks, the values of
$v_{osc}$ are in excellent accordance with the values of
$v_{det}$, which confirms that the primary peaks were caused by
the Josephson ac response, $\it{i.e.}$, integer Shapiro steps
\cite{Latyshev}. Frequencies corresponding to $V_{osc}$=38.9,
51.4, 54.4, and 60.8 mV are 0.67, 0.87, 0.94, and 1.06 THz,
respectively \cite{Frequency}. For the bias conditions of
$V_{osc}$=54.4 and 60.8 mV, the humps at $V_{max}$ in Fig. 3(b)
smear out, as the outward-shifting positions of the conductance
peaks overlap with those of $V_{max}$.

On the other hand, in Fig.3(b), the voltage positions of the
conductance sub-peaks turn out to be one half of $V^{prim}_{det}$.
The relation between the two kinds of peaks in terms of reduced
parameters is shown in Fig. 4(a). These half-integer Shapiro steps
are believed to be caused by phase locking of the second harmonics
of the Josephson oscillations on the primary frequencies
\cite{Likharev}. For the bias voltage of 100.8 mV in the
quasiparticle tunneling state of the oscillator stack, without the
vortex-lattice formation along the $c$ axis, no effective
vortex-flow-induced emission is expected. Thus, as in Fig. 3(b),
the response of the detector without any Shapiro-step peaks in
this bias is almost identical to the zero-bias curve represented
by the black curve.

\begin{figure}[t]
\begin{center}
\leavevmode
\includegraphics[width=1\linewidth]{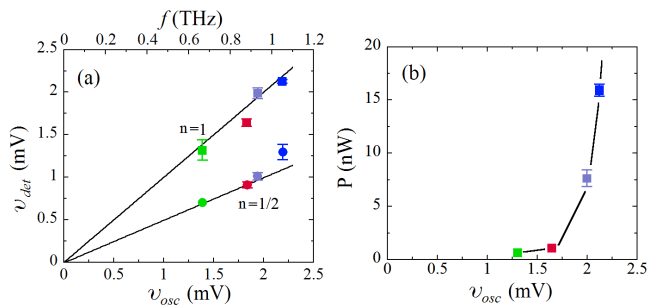}
\caption{(color online) (a) Primary ($n$=1) and sub-harmonic
($n$=1/2) conductance-peak voltages in the detector stack,
corresponding to the integer and the half-integer Shapiro steps,
respectively, with the emission frequencies denoted in the upper
horizontal axis. (b) Estimated power irradiated on the detector
stack for each bias condition in the oscillator stack.}
\end{center}
\end{figure}

The irradiation power, $P$, is estimated from the approximate
relation \cite{Power}, $P$$\sim$$\Delta
I_1^2V_1\Omega^2\beta_c/2I_c$, where $\Delta I_1$ is the height of
the Shapiro current step, $V_1$ is the voltage value of the
primary peak position in $N$-stacked junctions,
$\Omega^2\beta_c$=$(\pi h/e)(\epsilon f^2/tJ_c)$, $\Omega$=$f/f_c$
is the irradiated-wave frequency normalized by the characteristic
frequency $f_c$=$2eI_cR_n/h$, $\beta_c$ is the McCumber parameter,
$\epsilon$ is the dielectric constant, $t$(=1.2 nm) is the spacing
between neighboring CuO$_2$ layers, and $I_c$ ($J_c$) is the
tunneling critical current (density). The value of $\Delta I_1$ is
estimated using the relation $\Delta I_1$=$\Delta
V_1\frac{dI}{dV}|_{prim}$, where $\Delta V_1$ is the voltage width
of the primary peak \cite{Latyshev} estimated at the base level of
each curve in Fig. 3(b). Fig. 4(b) illustrates that the estimated
power irradiated onto the detector stack increases rapidly for
lower-index modes; $P$=0.69$\pm$0.14, 1.01$\pm$0.16, 7.5$\pm$0.7,
and 15.9$\pm$0.5 nW for the biases of $v_{osc}$=1.39, 1.84, 1.94,
and 2.17 meV, respectively. The emission power from the SJP modes
is indeed theoretically predicted to grow for lower-index modes as
the emitted waves become more phase-coherent and reaches a maximum
for the completely in-phase mode corresponding to $n$=1
\cite{Machida}. One also notices that the two bias conditions
$V_{osc}$=51.4 and 54.4 mV in Fig. 3(a), although in the same
branch, result in a large difference in the estimated irradiation
power as in Fig. 4(b). This takes place because, as the vortex
motion approaches the resonance condition in a branch, the energy
fed by the bias is consumed more for an emission rather than
speeding up the vortex lattice. This observed emission from the
vortex-flow region well justifies the excitation picture of the
SJP modes themselves. A more effective impedance matching scheme
between the oscillator and the detector stacks by combining proper
antennas in the system is expected to generate clearer Shapiro
steps with higher emission power.

In the presence of Josephson vortices a Josephson junction itself
is bound to exhibit a distributed Shapiro-step response even to
the monochromatic irradiation \cite{Latyshev}. Thus, the finite
Shapiro-step peak width in Fig. 3(b) does not necessarily
represent the frequency distribution in the emitted waves, which
can be determined by the heterodyne detection instead
\cite{Ustinov,Batov}.

The vortex-flow terahertz generation in stacked Bi-2212 IJJs has
the great advantage of continuous and frequency-tunable nature of
the generated waves. The coherency of the terahertz emission from
the rectangular vortex-lattice configuration is another advantage
of vortex-flow technique over, for instance, the radiation in a
quasiparticle-tunneling region where the phase coherence is
lacking between radiations from adjacent junctions. The direct
detection of emission in this study provides an additional noble
scheme of developing solid-state terahertz emission sources
\cite{Kadowaki} that eventually helps bridge the terahertz gap.

This work was supported by the National Research Laboratory
program administrated by Korea Science and Engineering Foundation
(KOSEF). This paper was also supported by POSTECH Core Research
Program.


\begin{thebibliography}{00}
\bibitem{Frequency} The frequency is estimated based on the frequency-voltage (per
junction) conversion relation, $f$=2e$V_{osc}$/$Nh$ (=483.6
GHz/mV$\times$$V_{osc}$/$N$).
\bibitem{Power} For the power formula, we used $P$=$I_{ac}^2R_{in}/2$,
where $I_{ac}$ is the ac current amplitude of irradiated waves,
$R_{in}$(=$N/2\pi fC)$ is the input impedance and $C$ is the
capacitance of a junction \cite{Rother}. Since the irradiated
power is expected to be low enough in this study the height of the
current step varies in linear proportion to the irradiated power,
resulting in $\Delta I_1\approx I_{ac}/\beta_c\Omega^2$.
\end{thebibliography}
\end{document}